\documentclass[a4paper,11pt]{article}
\pdfoutput=1 
\usepackage{jinstpub} 
\usepackage{lineno}

\usepackage{siunitx}
\usepackage{subcaption}
\usepackage{graphicx}

\title{\boldmath{Employing infrared microscopy (IRM) in combination with a pre-trained neural network to visualise and analyse the defect distribution in Cadmium Telluride crystals}\\ \vspace{1em}\small\textit
{
This is the Accepted Manuscript version of an article accepted for publication in \textit{Journal of Instrumentation}. IOP Publishing Ltd is not responsible for any errors or omissions in this version of the manuscript or any version derived from it. The Version of Record is available online at \url{https://doi.org/10.1088/1748-0221/16/08/P08044}.}
 }

\author[a,1]{S.~Kirschenmann,\note{Corresponding author.}}
\author[a]{S.~Bharthuar,}
\author[a]{E.~Br\"{u}cken,}
\author[a,b]{M.~Golovleva,}
\author[a]{A.~G\"{a}dda,}
\author[a]{M.~Kalliokoski,}
\author[a,b]{P.~Luukka,}
\author[a,c]{J.~Ott,}
\author[d, a]{and A.~Winkler}

\affiliation[a]{Helsinki Institute of Physics,\\Gustaf H\"{a}llstr\"{o}min katu 2, FI-00014 University of Helsinki, Finland}
\affiliation[b]{Lappeenranta University of Technology,\\Skinnarilankatu 34, FI-53850 Lappeenranta, Finland}
\affiliation[c]{Aalto University,\\Department of Electronics and Nanoengineering, Tietotie 3, FI-02150 Espoo, Finland}
\affiliation[d]{Detection Technology Plc,\\A Grid, Otakaari 5a, 02150 Espoo, Finland}

\emailAdd{stefanie.kirschenmann@helsinki.fi}

\abstract{While Cadmium Telluride (CdTe) excels in terms of photon radiation absorption properties and outperforms silicon (Si) in this respect, the crystal growth, characterization and processing into a radiation detector is much more complicated. Additionally, large concentrations of extended crystallographic defects, such as grain boundaries, twins, and tellurium (Te) inclusions, vary from crystal to crystal and can reduce the spectroscopic performance of the processed detector.
A quality assessment of the material prior to the complex fabrication process is therefore crucial. To locate the Te-defects, we scan the crystals with infrared microscopy (IRM) in different layers, obtaining a 3D view of the defect distribution. This provides us with important information on the defect density and locations of Te inclusions, and thus a handle to assess the quality of the material. For the classification of defects in the large amount of IRM image data, a convolutional neural network is employed. From the post-processed and analysed IRM data, 3D defect maps of the CdTe crystals are created, which make different patterns of defect agglomerations inside the crystals visible. In total, more than 100 crystals were scanned with the current IRM setup. In this paper, we compare two crystal batches, each consisting of 12 samples. We find significant differences in the defect distributions of the crystals.
}


\begin{document}
\maketitle
\flushbottom

\section{Introduction}
\label{sec:intro}

The development of medical devices further maximising the diagnostic and treatment success while minimising the radiation exposure of healthy tissue has been an ongoing effort for many years. Improved medical devices and technologies are crucial in the fight against cancer.\\ A novel direct detection system based on photon counting (PC), which registers the energies of the incoming photons in every pixel of the radiation detector as described in~\cite{Brucken2020}, would mean a significant advancement. For such a system a detector material is needed which can cope with high fluence rates, has a good stopping power, allows for a fast readout and energy discrimination, and is PC-capable.
Being a high-Z material with excellent photon radiation absorption properties, CdTe is a semiconductor which can outperform silicon, especially in clinically relevant energy regions of ionizing radiation. Furthermore, CdTe-based detectors can be operated at room temperature due to the wide band gap of 1.44\,eV~\cite{DelSordo2009}. However, CdTe ingots are difficult to grow and complicated to process~\cite{Gadda2017, Winkler2019}.
Moreover, the quality of the CdTe material can differ between the raw crystals: CdTe consists of the two elements Cadmium and Telluride organized in a crystalline zincblende structure~\cite{zincblende}, but can include deviations from this structure, e.g.~monocrystalline tellurium (Te) inclusions. These inclusions (or \textit{defects}) can deteriorate the performance of the processed detector, as they can act as traps for the charge carriers and therefore lead to a loss in charge collection efficiency (CCE)~\cite{DelSordo2009,Winkler2019}. For these reasons it is important to estimate the quality of the CdTe material before starting the demanding processing into a radiation detector (described in~\cite{Gadda2017}). \\
Previous research shows that the amount of defects in CdTe as well as their sizes can play an important role in estimating the quality of the material and the performance of the detector~\cite{Tepper2008, Bolotnikov2007, Bolotnikov2013}: In~\cite{Tepper2008} a strong correlation between the volume fraction of Te defects and carrier trapping time is found and it is pointed out that defect density measurements could be used as a screening tool for material selection. The effect of defects on the electron transport and charge loss in a radiation detector have been modelled in~\cite{Bolotnikov2007} and their impact on the detector's performance predicted in~\cite{Bolotnikov2013}. Obtaining information on the defect distribution inside the CdTe crystals is therefore an important part in the quality assessment.\\
Infrared microscopy (IRM) is a non-invasive method that can be used to study defects (Te inclusions) not only on the surface, but also in different depths inside the CdTe material, which is a limit for other methods such as X-ray response maps, a method used e.g. in~\cite{Bolotnikov2013}. As the compound material CdTe is mostly transparent in the near-infrared light~\cite{Winkler2011, CdTeTransmission}, but the Tellurium inclusions are opaque for light in this region, the defects become visible in the IRM images.
In~\cite{Shiraki2010}, IRM was employed to investigate the behaviour of Te inclusions for different growth rates. In~\cite{Bolotnikov2008} and~\cite{Roy2011}, infrared microscopy was used in combination with an iterative detection algorithm to obtain distributions of the concentration and size for THM (Travelling Heater Method) grown CdZnTe samples. \\
In order to obtain defect data for a larger amount of crystals, a higher area-coverage, and different depths, an alternative to using an iterative algorithm is presented in this paper: A convolutional neural network (NN) trained on finding defects in IRM images obtained with our setup is used for the defect-detection in the huge amount of IRM images. We then analyse and post-process the NN-output and show how the post-processed data can be used to create three-dimensional defect maps of CdTe crystals. The combination of NN and post-processing techniques makes the comparison of defect data between different crystals possible on a larger scale. We observe and discuss differences between crystals and crystal batches, focusing on the comparison of two batches of crystals, each consisting of 12 CdTe crystals, respectively.

\section{Materials and Methods}
\subsection{Quality estimation of CdTe crystals using infrared microscopy (IRM)}
An infrared microscopy setup as introduced in~\cite{Winkler2011} and described in~\cite{Gadda2017} is used to obtain images of the surface as well as of the inside of a CdTe crystal. A white light source (quartz-halogen) acts as the infrared source: The CdTe crystal itself omits wavelengths below the IR transparency and the silicon of the camera sensor the longer wavelengths, due to its very low quantum efficiency for wavelength above 1\,$\mathrm{\mu m}$~\cite{Winkler2011}. The sample holder is mounted on an x-y-z stage (\texttt{Steinmeyer MP63-25-DC-R}). By moving the stage along the optical axis, different layers inside the crystal get into the focus of the optical system. Moving the sample in z-direction, the refractive index for CdTe has to be taken into account, which is between $\approx$ 2.8-2.9 in the relevant region~\cite{RI_CdTe}: A movement of $\approx$ 345$\mathrm{\mu m}$ in z-direction already corresponds to the light passing the whole geometric length of a sample of 1\,mm. The z-movements in the measurements are adjusted accordingly using the compressed values, thus a stepping value of $100\mathrm{\mu m}$ in z-direction corresponds in reality to a physical length of almost $300\mathrm{\mu m}$. For better readability and to avoid confusion between the different lengths, we will in this paper refer to the z-dimensions by their more common physical length. The crystal is moved in a predefined manner automatically along the three dimensions, taking an IRM image at every position. 
The resulting IRM images show dark objects of different sizes, which can be identified as the above-mentioned defects. In principle, defects down to a minimum size of about 1\,$\mathrm{\mu m}$ can be resolved with near-infrared light~\cite{Winkler2019}. However, defects with a size of less than 2-3\,$\mathrm{\mu m}$ cannot be clearly identified in the IRM images.
The THM-grown crystals scanned for this study have dimensions of $10\times10\times1$\,\si{mm^3} and were manufactured by Acrorad Co., Ltd., Japan. Other thicknesses between 0.5 and 2\,mm were measured in smaller amounts as well, but for comparability CdTe crystals with a common thickness of 1\,mm were studied in this paper. The field of view (FOV) of the camera\footnote{CMEX-1300x, with a pixel size of $3.6\,\mathrm{\mu m}\times 3.6\,\mathrm{\mu m}$ and image dimensions of $1280\times 1024$ pixel.} in the current setup is $\approx 200\,\mathrm{\mu m}\times160\,\mathrm{\mu m}$, thus, to facilitate a complete layer scan, over 3000 images are taken. The total amount of images scales with the number of layers and crystals scanned. Every image taken accounts for a unit volume of $\approx(200\times160\times150)\,\mathrm{\mu m}^3=4.8\cdot10^6\mathrm{\mu m}^3$, approximating the z-dimension of the volume with $150\,\mathrm{\mu m}$. Whereas the estimated focal depth along the optical axis -- the distance from the focal plane in which an object is still seen in complete focus -- is $\approx$ 10\,$\mathrm{\mu m}$~\cite{Winkler2019}, we can still see unfocused objects from up to about $75\,\mathrm{\mu m}$ distance from the layer in focus. This factor for the visibility in depth is larger than the assumed $50\,\mathrm{\mu m}$ in~\cite{Winkler2019}, and was obtained as an intermediate value studying the visibility of different defects along the optical axis (cf. section \ref{sec:post-proc}). For the analysis, we want to include these out-of-focus objects, to get a better estimate on the total number of defects. By dividing the number of defects per image by the unit volume, we can calculate the defect density in the respective area. 
To evaluate the defect density of the whole crystal, the defects in all the images need to be counted and classified. In order to make this quality estimation by defect density measurements a feasible step in the detector development process, some automation is required, as manual classification would be too time-intensive. This is why we introduce a convolutional neural network to facilitate the detection and classification of defects found in the IRM images. In this paper, we refer to the different crystals by their batch number (1 or 2) and a supplemental crystal index (1 to 12).

\subsection{Neural network evaluation}
\label{sec:NN}
As the IRM scanning provides large quantities of image data, supervised machine learning is a suitable technique to work with for finding patterns or objects (in our case: \textit{Te inclusions, defects}) in these images. Manually classifying a certain amount of already existing data (IRM images), a training data set can be created and used to train a NN. The trained model can then be applied on new IRM data.\\
The Ohtu NN\cite{Ohtu} used in this study has been explicitly developed for the defect detection in IRM images. Apart from applying it exemplarily on two crystals with low resolution without any post-processing in \cite{Winkler2019}, it has however not been applied before on new crystal data. The NN has been studied in~\cite{ErsalanThesis}, where it was concluded that it is already sufficiently trained. However, limits in the accuracy of the NN exist, which make a post-processing of the neural network results necessary. The developed post-processing procedure and improvements obtained will be discussed in the results sections. Furthermore, suggestions for an improved NN version will be derived.
The Ohtu NN employs \texttt{Keras}~\cite{chollet2015keras} on top of a \texttt{Tensorflow}~\cite{tensorflow2015-whitepaper} backend. The detection and classification of the defects in the IRM images is facilitated by building upon the Keras implementation of RetinaNet (\texttt{keras-retinanet 0.2}~\cite{keras-retinanet}). The object detection with \texttt{RetinaNet} is described further in~\cite{Lin2017}: RetinaNet introduces the \texttt{Focal Loss} function as a loss and its architecture includes a \texttt{Feature Pyramid Network} as a backbone on top of a feedforward \texttt{ResNet}~\cite{He2015} architecture.\\
A \texttt{MATLAB}~\cite{MATLAB:2018}-based tagging tool~\cite{Ohtu} facilitates the training of the Ohtu neural network: Via a GUI, defects in IRM images are tagged, recording the position, chosen defect shape category (\textit{label}) and bounding box coordinates into a data file. We use the latest Ohtu predictive inference model version, which was trained on about 18000 IRM images, including one random image augmentation (horizontal and vertical flips, rotations) per image. The IRM images for the training set were taken with the same experimental setup and camera used for this study. The predictive model returns the bounding box coordinates, a label and a label score for every defect found. In addition, it creates image copies with annotated bounding boxes (cf. Fig.~\ref{fig1}) around the detected defects~\cite{Ohtu}.

\section{Analysis and Results}
\subsection{Post-processing of the IRM data}
\subsection*{IRM images}

\begin{figure*}[htb]
  \centering
  \begin{subfigure}[b]{0.45\linewidth}
    \centering
     \includegraphics[width=\linewidth]{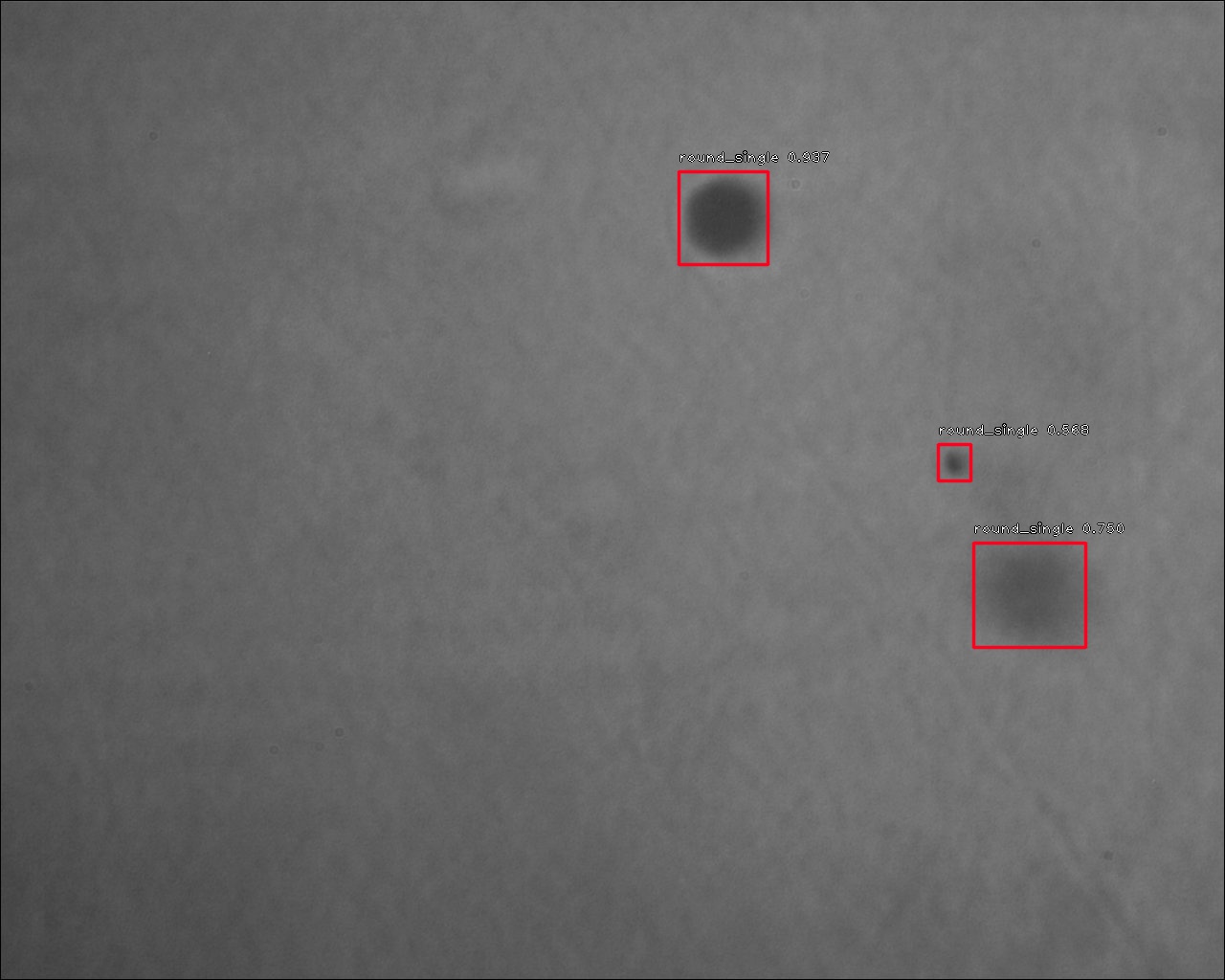}
    \caption{}
  \end{subfigure}
  \hspace{1em}
  \begin{subfigure}[b]{0.45\linewidth}
    \centering
    \includegraphics[width=\linewidth]{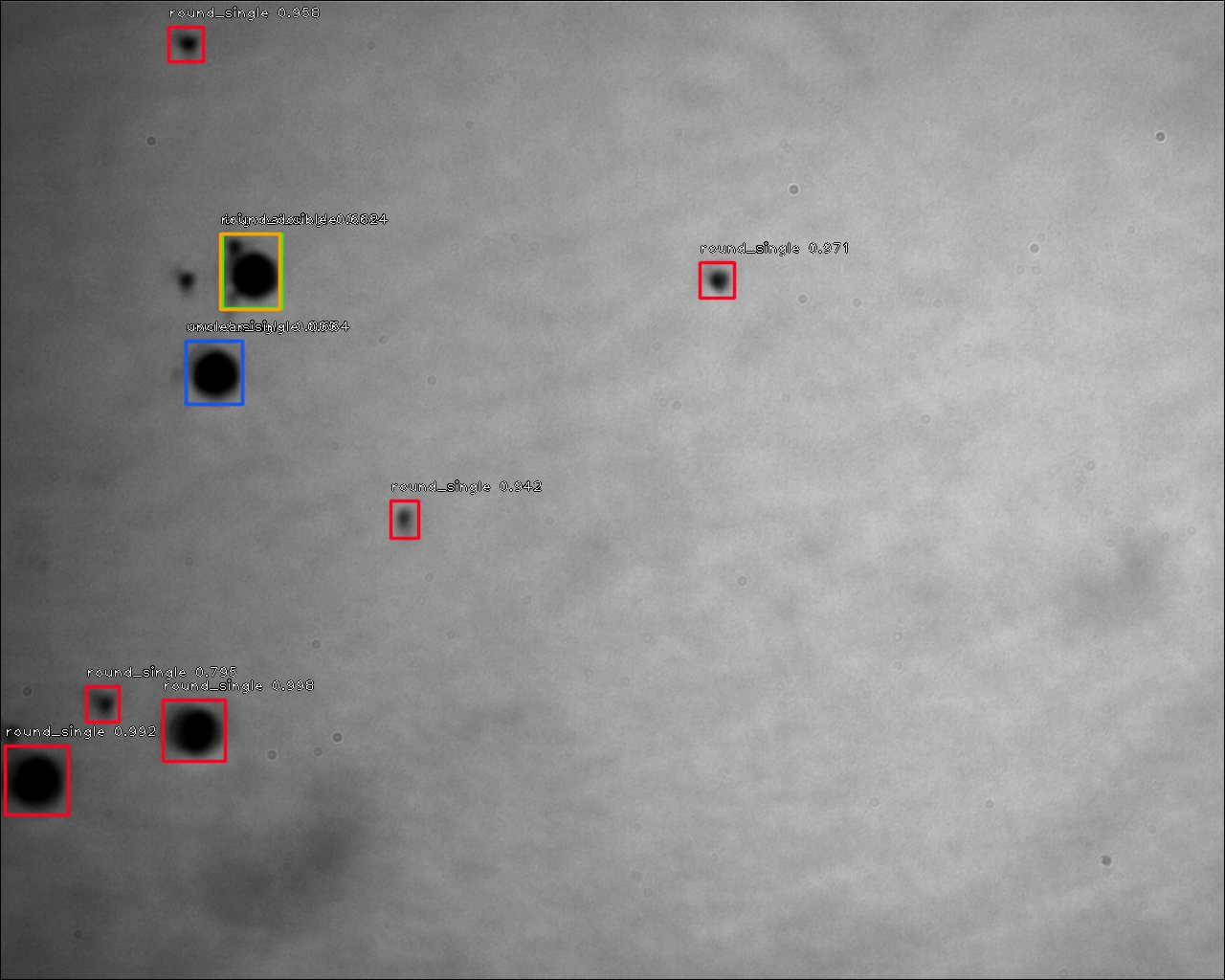}
    \caption{}
  \end{subfigure}
  \caption{Example of IRM images evaluated by the neural net (FOV 200\,$\mu$m x 160\,$\mu$m),  b) showing more visible objects than a). Blue denoting category \textit{unclear single}, red category \textit{round single}, orange category \textit{round double} and green category \textit{trigonal single}. Two defects in the upper left corner (next to the unmarked one) are marked twice.}
  \label{fig1}
\end{figure*}

In Fig.~\ref{fig1} two exemplary IRM images with objects identified by the NN are shown. For Fig.~\ref{fig1}a the NN marks the three visible defects that would have been also tagged manually. The color of the bounding box refers to different categories (\textit{red}: round single, \textit{blue}: unclear single). The defect appearing more blurry is further away from the focal point than the other two identified defects.

Fig.~\ref{fig1}b shows another image obtained via IRM. Contrary to the area in Fig.~\ref{fig1}a, this area shows more objects, some clearly identifiable as tellurium inclusions and most of them in focus. At least one defect that can be seen by eye is not marked at all and two objects are tagged twice, classifying them into two categories simultaneously. This double counting will be addressed with a cut as described in the next section.

\subsection*{Performance of the employed NN}
\label{sec:performance}
The total number of identified defects in the evaluated set of images gives an estimate for the overall defect density, and the defect coordinates $[$x1, y1, x2, y2$]$ of the bounding boxes around the defects lets us determine the approximate size and position of the defects inside the crystal. For the NN results, we reduced the before-mentioned duplicate counts for the same defect (cf. Fig.~\ref{fig1}) in the same image with the following cut:  The center coordinates of every found object were calculated and objects close to each other (\textit{here}: 10 pixels apart in both x- and y-coordinates) were compared with respect to their category and label score. If one of the duplicates was of category \textit{unclear single} it was rejected, as the label \textit{unclear single} is the most unspecific. Otherwise, the object with the lesser score was rejected. The filter was applied to all the IRM evaluated data and led to a reduction of the defect count per crystal of about 4-8\,\%. It was further found out that the objects of the category \textit{square single} can be rejected as those are image artefacts on the edges of the crystal, which led to another reduction of 4-12\,\%. 
Most of the objects are either classified \textit{unclear single} or \textit{round single}. The category \textit{unclear single} is relatively unspecific, containing objects whose shape cannot be clearly labeled as \textit{hexagonal} or \textit{round}. This leads to a high inaccuracy concerning the category label.
However, a deeper discussion on the defect shape is not necessary for the presented results as we focus on the defect density.\\
To evaluate the performance of our neural network, in particular concerning the suitability for evaluating the local defect densities, we compared the predictions of the NN for one crystal surface to the result of manually labeling the same image set. Employing the above-mentioned cut leads in this particular case to an overall count reduction of about 11\,\%.
\begin{figure}[htb]
	\centering
\includegraphics[width=0.5\textwidth]{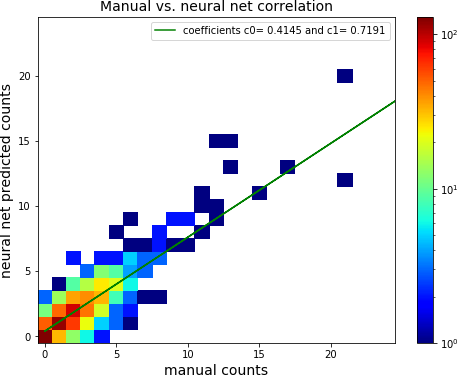}
\caption{Correlation between number of defects identified by the NN and by manual labelling.}
\label{correlation:man-nn}
\end{figure}
The correlation between the defect counts obtained from the NN prediction and from manually labeling is shown in Fig.~\ref{correlation:man-nn}. The linear correlation is clearly visible. We can find that the correlation between manually labelled counts and NN predicted counts is best fit by a linear relation $$y(x)=c_0 + c_1\cdot x,$$ with $c_0=0.415\pm0.045$,  $c_1=0.719\pm0.013$ and $R^2=0.764$.
After the cuts, the NN predicted counts are about 10\% less than the manual counts. This can be mostly explained by the missing identification of defects between 2 and 5\,$\mathrm{\mu m}$. These were included in the manual labeling, but the original Ohtu NN was not trained in this region and was used as is. However, with respect to defect finding for sizes above 5\,$\mathrm{\mu m}$, the NN performs reasonably well and enables pattern-finding.

\begin{figure}[htb]
	\centering
\includegraphics[width=0.95\textwidth]{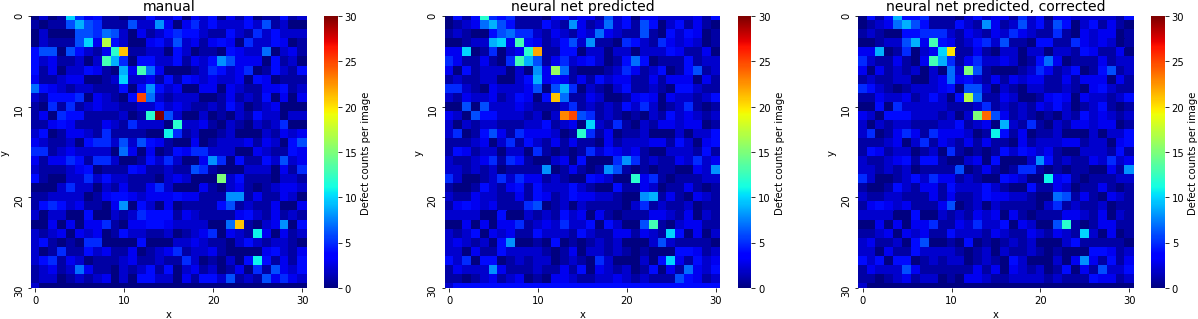}
\caption{Defect map of a crystal layer, showing the defect counts per position. \textit{Left}: manually labelled, \textit{center}: NN-evaluated \textit{right}: NN-evaluated and corrected.}
\label{layer1:man-nn}
\end{figure}

As an example, a defect map was created from the compared image set, both for the manually-labelled counts as well as for the NN predicted counts. The left plot in Fig.~\ref{layer1:man-nn} shows the manual counts, the center and right plot the defect maps from the NN counts, before and after the correction. As can be seen in Fig.~\ref{layer1:man-nn}, the same line-pattern is still visible in all three plots.

\subsection{Analysis of post-processed IRM data}\label{sec:post-proc}
\paragraph*{Defect maps}
\begin{figure*}[htb]
  \centering
  \begin{subfigure}[b]{0.45\linewidth}
    \centering
    \includegraphics[width=\linewidth]{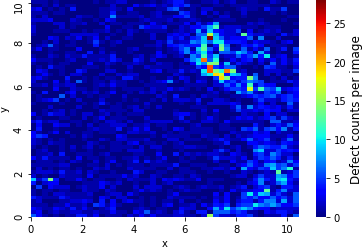}
    \caption{}
  \end{subfigure}
  \hspace{1em}
  \begin{subfigure}[b]{0.45\linewidth}
    \centering
    \includegraphics[width=\linewidth]{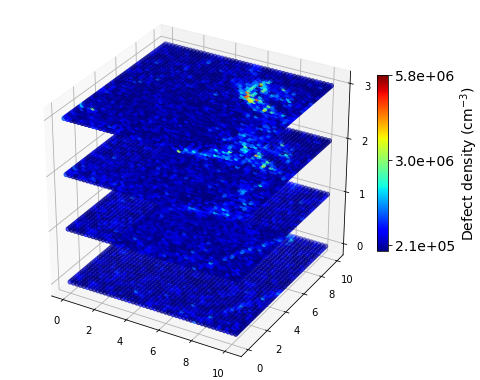}
    \caption{}
  \end{subfigure}
  \caption{a) Defect map of the back surface of a CdTe crystal (Batch 2-05): every rectangle represents the defect count of one image taken with IRM, in total 51x64 images. Dimensions on the x- and y-axis are in mm. b) 3D defect map showing the defect density in 4 different layers (0: front, 1: 45\,$\mathrm{\mu m}$, 2: 955\,$\mathrm{\mu m}$, 3: back).}
  \label{defectmap}
\end{figure*}
From the output data we create defect maps of crystal layers, showing how many defects per image have been counted. An exemplary surface defect map for a crystal (Batch 2-05) is shown in Fig.~\ref{defectmap}. For every image position the respective defect count inside that image is plotted, for 51x64 images in total.
This can make accumulations of defects in the defect maps visible. By aligning the defect maps of different layers, we obtain a 3D-distribution of the defects inside the crystal (cf. Fig.~\ref{defectmap}b). For the back layer of this sample, we see areas of high local defect counts of up to 28 counts per image (cf. Fig.~\ref{defectmap} a), corresponding to defect densities of up to $5.8\times 10^6\,\mathrm{cm}^{-3}$ (cf. Fig.~\ref{defectmap} b, layer 3). 
The z-axis in the 3D plot refers to the different layers scanned inside this crystal. Layers 0 and 3 denote the front and back surface, whereas layer 1 and 2 are at a distance of 45\,$\mathrm{\mu m}$ from the respective surface. One can see a clear aggregation of defects in layers 3 and 2 of the sample. On layers 0 and 1 a line is visible which could show the position of a grain boundary. As the layers 0 and 1 (and 2 and 3) are so close to each other, we count many defects in both layers. This will be addressed in the next section.

\begin{figure}[htb]
	\centering
\includegraphics[width=0.5\textwidth]{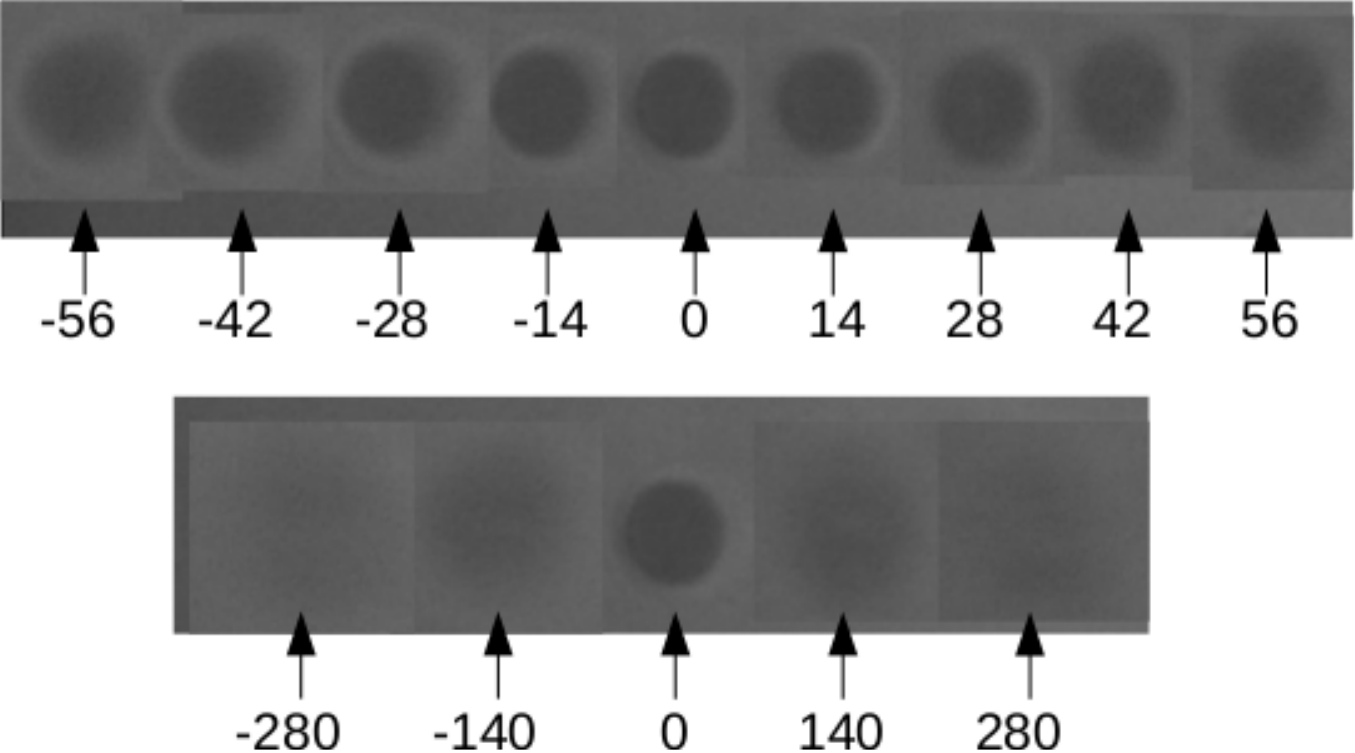}
\caption{The image shows how a defect of size $10\,\mathrm{\mu m}$ (in focus in the center) gets blurry while getting out of focus. \textit{upper part}: $\approx$ 14\,$\mathrm{\mu m}$ steps, \textit{lower part}: $\approx$ 140\,$\mathrm{\mu m}$ step.}
\label{out-of-focus}
\end{figure}

\paragraph*{Defect-size distribution}
The bounding-box coordinates inferred from the data let us determine an approximate diameter (size) of an object by taking the mean of its x and y dimensions. The following errors have to be considered:
We assign a systematic uncertainty of 2.5\,\% on the defect size related to the FOV dimensions. 
Furthermore, the error on the size for objects detected by the NN depends on how much they are in focus. 
Fig.~\ref{out-of-focus} shows one detected defect of about 10\,$\mathrm{\mu m}$ diameter and follows it through several adjacent layers. The upper part of Fig.~\ref{out-of-focus} shows smaller z-steps ($\approx 14\,\mathrm{\mu m}$) between the layers looked at, the lower one z-steps of about 140\,$\mathrm{\mu m}$. For an object out of focus, the NN includes the blurry corona of the object in the bounding box. This overestimates the size of the objects by up to 40\,\% for larger and up to 70\,\% for smaller defects. This error on the size is smaller for objects of the NN-labeled category \textit{round single}, as these are mostly quite close to the focus. Therefore, when studying the defect size distribution, it is more appropriate to only include the round single category, whereas for the defect maps we would want to include all detected defects. 

\begin{figure*}[htb]
  \centering
  \begin{subfigure}{0.45\linewidth}
    \centering
    \includegraphics[width=\linewidth]{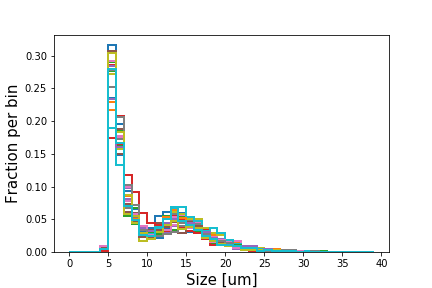}
    \caption{ }
  \end{subfigure}
  \hspace{1em}
  \begin{subfigure}{0.45\linewidth}
    \centering
    \includegraphics[width=\linewidth]{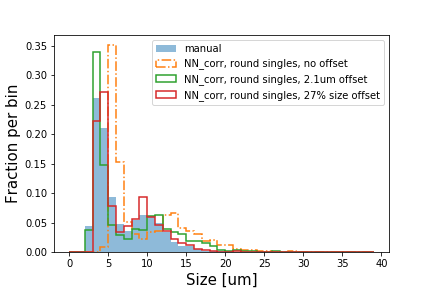}
    \caption{ }
  \end{subfigure}
  \caption{a) Defect size distribution of different crystals (Batch 1) for defects classified as \textit{round single}. b) Defect size distribution for the manually labeled crystal layer compared to the round single objects from the NN-evaluation (cf. Fig.~\ref{layer1:man-nn}), applying two different offsets.}
  \label{cmpCrys}
\end{figure*}

Fig.~\ref{cmpCrys}a shows the defect size distribution for crystals of batch 1 obtained from the NN evaluated data, taking only into account the defects of category \textit{round single}. The highest peak in the distribution lies for all crystals at smaller defect sizes, between 5 and 7$\mathrm{\mu m}$. No conclusions can be drawn for the fraction of defects with size below $5 \,\mathrm{\mu m}$ as these are not clearly identifiable with the current NN.

Fig.~\ref{cmpCrys}b shows a comparison between the size distribution obtained from the manually labeled data (cf. Fig.~\ref{layer1:man-nn} in section \ref{sec:performance}) and NN labeled data for the same image set. We see a deviation between the manually labeled data (blue) and the distribution obtained from the whole NN dataset (orange, dotted). For the manual data a clearer second peak around 10\,$\mathrm{\mu m}$ is observed and also a shift of the main peak towards lower sizes.
When manually labeling it was decided to restrict the bounding boxes of out-of-focus defects to the core of the defect, approximately equivalent in size to an in-focus object and excluding the corona, in order to get a better estimate on the actual defect size. 
Additionally, also smaller defects (down to $2\,\mathrm{\mu m}$) were tagged. This explains the shift towards lower sizes and could imply an even larger fraction of defects below the detection limit. 
We studied the $\chi^2$ distance between the histograms for different offsets and see that an offset of 27~\% on the size as calculated from the NN-output or alternatively an offset of $2.1\,\mathrm{\mu m}$ perform best. The green and red lines in Fig.~\ref{cmpCrys}b show the distributions including the respective offset that are in good agreement with the manually labeled distribution. The offset is due to the larger bounding boxes in the NN detection because of a small corona effect (cf. Fig.~\ref{out-of-focus}) in comparison with the tighter bounding boxes in the manual labeling.
Even with these size uncertainties, both the distributions in Fig.~\ref{cmpCrys}b as well as the distributions in Fig.~\ref{cmpCrys}a show clearly, that apart from the highest peak at smaller values, we get a second maximum at higher defect sizes. 
This second maximum is approximated to be in the region of 10 to 15\,$\mathrm{\mu m}$, if we take into account the shift due to the overestimation of defect sizes by the NN. The distributions are in general agreement with earlier research \cite{Roy2011, Tepper2008}: We see a higher contribution of smaller defects than of larger ones and we either have two main peaks -- one at smaller, one at medium to larger sizes -- or
an almost exponential decrease from smaller to larger sizes.

\paragraph*{IRM depth scan}
\begin{figure}[htb]
	\centering
\includegraphics[width=0.6\textwidth]{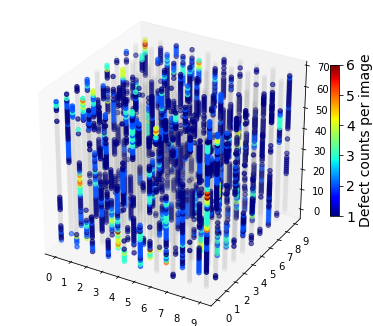}
\caption{Defect map of the lower left corner through the whole crystal.}
\label{lowerleft}
\end{figure}
Fig.~\ref{lowerleft} shows a 3D defect map of a corner of one CdTe crystal ($2\times1.6\times1\,\mathrm{mm}^3$), for which the z-step between adjacent layers has been reduced to 15\,$\mathrm{\mu m}$, 68 layers in total. 
This scan provides more insight on the correlation between detection and focus.
One can see how some defects (or accumulation of defects) extend over several adjacent layers. To elaborate, we simplify a typical medium sized Te defect as a sphere with a diameter of 15\,$\mathrm{\mu m}$ with its center in the focus plane. The defect will be visible at this certain depth in complete focus, as well as almost in focus in the two adjacent layers.  Due to the geometry of the optical system, the defect is still visible -- although blurred -- for a certain distance to the focus plane (cf. Fig.~\ref{out-of-focus}). The NN detects this example defect for $\approx 200 \,\mathrm{\mu m}$, which is about a fifth of the depth of the 1\,mm thick crystal. As larger defects extend more into the z direction, they will be visible throughout a larger z range than smaller ones. Taking this into account, we approximate the visibility in depth with a medium value of $150\,\mathrm{\mu m}$.
Our current scanning procedure of a 1\,mm crystal includes scanning of both crystal surfaces as well as two layers in between, with equidistant layers ($\approx 330\, \mathrm{\mu m}$ apart)\footnote{Scanning procedure for most of the samples of Batch 2, cf. section \ref{cmp-batches}.}. These layers are far enough apart to prevent double-counting of defects in adjacent layers. This reduces duplicates, but lowers the total defect count efficiency per crystal to about 50\,\%. A six layer scan with equidistant layers of 150\,$\mathrm{\mu m}$ distance starting at 75\,$\mathrm{\mu m}$ depth, would increase the detection efficiency. However, we would get significantly more double counts in such a scan for larger objects. This can be seen in the next section for Batch 1, where the scanning procedure included three equidistant layers of 150\,$\mathrm{\mu m}$ distance.

\subsection{Comparison of different CdTe crystals and crystal batches}
\label{cmp-batches}

With the above-described machine-learning, analysis and visualisation techniques, we can now compare different batches of CdTe crystals with respect to their defect density. This makes an easy first comparison between the crystals possible and offers a possibility for sorting before processing.

\begin{figure}[htb]
	\centering
\includegraphics[width=0.9\textwidth]{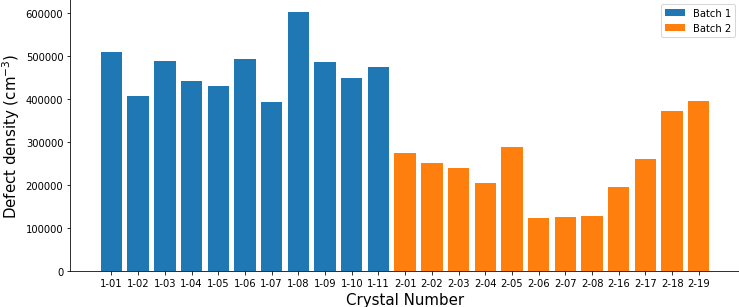}
\caption{Comparison of the mean defect density values of the crystals of batch 1 and batch 2. The crystals of batch 1 show in general a higher defect density than the crystals of batch 2.}
\label{defectdensity-cmp}
\end{figure}
\begin{figure}[htb]
  \centering
\includegraphics[width=0.95\linewidth]{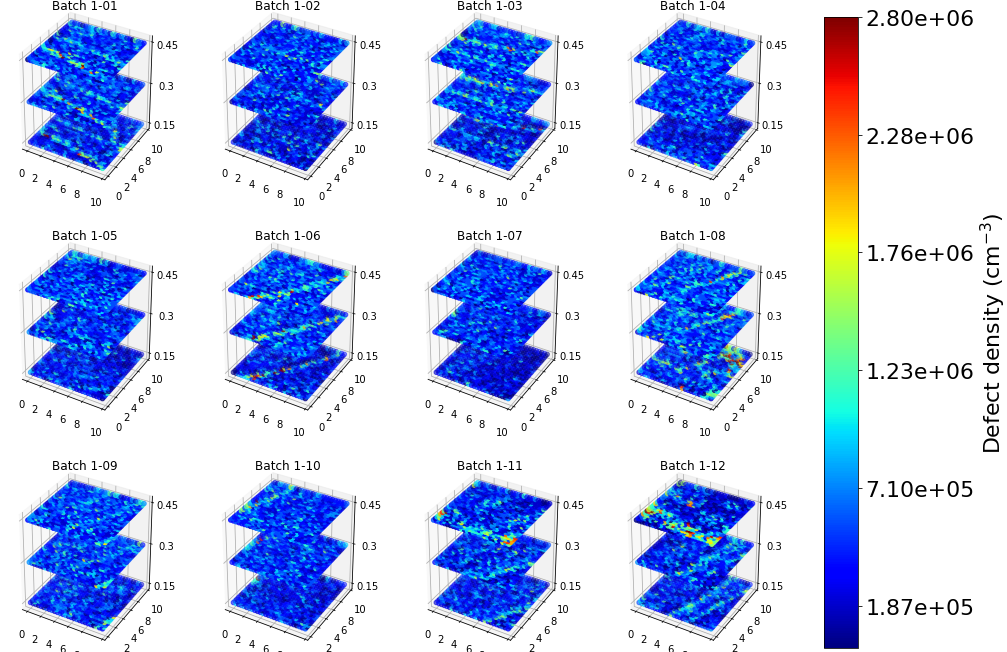}
\caption{3D defect plots for crystals from Batch 1, dimensions in mm: Structures are clearly visible in some crystals.}
\label{3d-cmp}
\end{figure}
  
\begin{figure*}[htb]
  \centering
  \begin{subfigure}{\linewidth}
    \centering
    \includegraphics[width=0.95\linewidth]{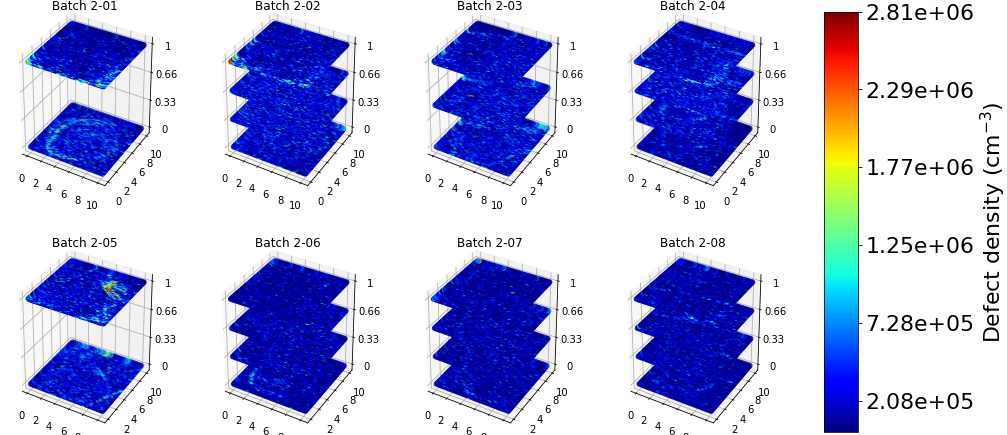}
  \end{subfigure}
  \begin{subfigure}{\linewidth}
    \centering
    \includegraphics[width=0.95\linewidth]{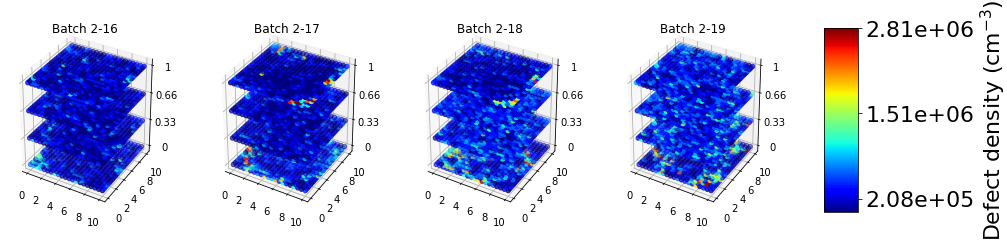}
  \end{subfigure}
  \caption{3D defect plots for crystals from Batch 2, dimensions in mm: Batch 1 shows an overall higher defect density than Batch 2 (cf. Fig.~\ref{3d-cmp}).}
  \label{3d-cmp-2}
  \end{figure*}

Fig.~\ref{defectdensity-cmp} gives a general overview of the mean defect density in the two compared batches and Fig.~\ref{3d-cmp} and \ref{3d-cmp-2} show 3D-defect density maps for the 2 different batches. In Fig.~\ref{defectdensity-cmp} it is already clearly visible, that batch 1 shows in general a higher defect density than batch 2, but that in-batch differences exist as well. However, this bar plot reduces the defect density to a single number; 3D maps of the defect distribution can provide further information on local areas of higher defect density.

\begin{figure}[htb]
	\centering
\includegraphics[width=0.9\textwidth]{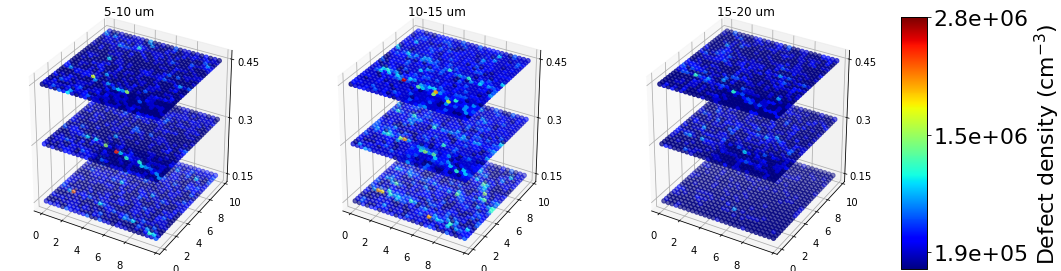}
\caption{3D defect maps for crystal Batch 1-01, showing the contribution of different size categories. \textit{Left}: defect size between 5 and 10\,$\mathrm{\mu m}$, \textit{center}: between 10 and 15\,$\mathrm{\mu m}$, \textit{right}: between 15 and 20\,$\mathrm{\mu m}$.}
\label{size-cuts}
\end{figure}

Some of the scans (Batch 1, Batch 2-16 to 19) were performed with a more coarse x- and y-step size. In this way not all smaller details can be identified, but the larger trends such as boundaries are still visible. This type of scan allows us to save time in the overall analysis.
In the 3D-defect density maps for Batch 1 (cf. Fig.~\ref{3d-cmp}), differences between crystals and layers are clearly visible. Following the categories for high, moderate and low defect density as defined in \cite{Winkler2019}, we can see here as well that the first batch (Fig.~\ref{3d-cmp}) shows a much higher defect density (between a mean value of 4.3 and 6.7$\times10^5\,cm^{-3}$)\footnote{to compare with the values in \cite{Winkler2019} we have to multiply these values by a factor of 3 to account for the smaller visibility in depth of 50\,$\mathrm{\mu m}$ assumed there} in all of the crystals than batch 2 (Fig.~\ref{3d-cmp-2}). Furthermore, the visualisation makes grain boundaries and areas of high defect density visible. Some lines are visible in two or even three layers (Batch 1-01, 1-03, 1-06, 1-08, 1-11, 1-12). As we have three equidistant layers of distance 150\,$\mathrm{\mu m}$ for batch 1, we will also get some double counts for large defects. However, as can be seen in Fig.~\ref{size-cuts} for the example of crystal Batch 1-01, smaller, as well as medium-sized defects are contributing to these boundaries. This confirms that the visibility of the line in all three layers is not just due to double-counting, but that the line extends over a larger area in the crystal.

For batch 2 (Fig.~\ref{3d-cmp-2}), most of the crystals were scanned in four equidistant layers (330\,$\mathrm{\mu m}$ apart). For Batch 2-01 and -05 we only show two layers, as for these two crystals pairs of close-by layers were scanned (cf. Fig.~\ref{defectmap}). Batch 2-03 shows three equidistant layers. The mean defect density for batch 2 is for most of the crystals between low and moderate (between 2.1$\times10^5$ and 4.3$\times10^5\,\mathrm{cm}^{-3}$). Even if the mean defect density of this batch is lower, one can still see different patterns in the defect accumulations. Crystals 2-01, 2-04, 2-06 as well as 2-17 and 2-19 show a circular artefact in at least one of the scanned layers and e.g crystal 2-08 shows a clear stripe (grain boundary). Furthermore, areas of large local defect density are visible in the different crystals (e.g. 2-05). 

\section{Conclusions}
A material assessment of bare CdTe crystals can be helpful to evaluate the quality of the material before the difficult and time-consuming processing of the crystals into radiation detectors for medical imaging. Defects (Tellurium inclusions) inside the CdTe material can trap charge carriers which leads to a reduction in the charge collection efficiency and can deteriorate the detector's performance. Therefore, it is beneficial to study the defect distribution beforehand.\\
In this study we introduced 3D infrared microscopy in combination with machine learning and data analysis techniques as a method to not only get an estimate of the overall (Te) defect density inside a CdTe crystal, but to also locate regions of higher defect densities. The application of machine learning methods -- in this study the employment of a neural network -- is essential to keep the time effort for this quality assurance step reasonable. 
Although the employed NN is not performing perfectly on a single defect basis, it was shown that it gives a good estimate on the quality of the material considering a whole set of images. Furthermore, it renders the comparison of several crystals possible with respect to their defect density: With 3D defect maps, patterns of higher defect density become visible for the different crystals. This lays the foundation for a more thorough study on the relation of these areas of higher local defect density and the performance of the detector. \\
For the presented findings, the defect shapes were not of high importance. However, an interesting follow-up study could be an in-depth study of different defect shapes, their sizes and their respective concentrations inside CdTe crystals. We already see with the current NN a second peak of medium-sized inclusions (estimated to be at 10-15\,$\mathrm{\mu m}$) in the defect distributions, which should be investigated further. To enable a more qualitative study of these larger inclusions and get a higher accuracy on the defect sizes we envision a retraining of the NN, with more clearly defined defect shape categories.\\
With regard to the IRM setup, technical improvements are considered as well. The current IRM setup is a prototype designed to line out the fundamental possibilities of such a system. We plan a reference IRM measurement setup, which could, with appropriate mechanics, optics, a stronger light source (e.g. laser) and a more sensitive camera, improve the image quality and allow for a study of crystals with higher thicknesses.
Further analysis and comparison of the IRM data with measurement results from transient current technique (TCT) and simulations with TCAD (Technology Computer Aided Design) are ongoing to study the impact of the different kind of patterns of higher local defect densities on the performance of the processed detectors.

\acknowledgments
This study was performed in the framework of the Academy of Finland project, number 314473, \textit{Multispectral photon-counting for medical imaging and beam characterization}, for which we would like to acknowledge the funding. S. Kirschenmann and S. Bharthuar thank the Magnus Ehrnrooth foundation for financial support. The measurements have been performed at the joint detector laboratory of the University of Helsinki and the Helsinki Institute of Physics and we are grateful to the staff for their support.

\bibliographystyle{JHEP}
\bibliography{CdTe_IRM_NN_2021_arxiv}

\end{document}